\documentclass[letterpaper, 10 pt, conference]{ieeeconf}  

\IEEEoverridecommandlockouts                              

\usepackage{graphics} 
\usepackage{amsmath} 

\usepackage{xcolor}

\usepackage{siunitx}
\usepackage{cite}

\newtheorem{remark}{Remark}

\title{\LARGE \bf
	Electrical neurostimulation for chronic pain:\\
	on selective relay of sensory neural activities in myelinated nerve fibers
}

\author{%
	Pierre Sacr\'{e}, Sridevi V. Sarma, Yun Guan, and William S. Anderson
	\thanks{P.~Sacr\'{e} and S.~V.~Sarma are with the Institute for Computational Medicine and the Department of Biomedical Engineering, The Johns Hopkins University, Baltimore, MD (p.sacre@jhu.edu, sree@jhu.edu).}%
	\thanks{Y.~Guan is with the Department of Anesthesiology/Critical Care Medicine, The Johns Hopkins University School of Medicine, Baltimore, MD (yguan1@jhmi.edu).}
	\thanks{W.~S.~Anderson is with the Institute for Computational Medicine and the Department of Neurosurgery, The Johns Hopkins University School of Medicine, Baltimore, MD (wanders5@jhmi.edu).}%
	}

\newcommand{\Iion}{I}

\newcommand{\eg}{for example,}

\graphicspath{{./figures/}{./}}

\begin{document}

\maketitle
\thispagestyle{empty}
\pagestyle{empty}

\begin{abstract}
Chronic pain affects about 100 million adults in the US. Despite their great need, neuropharmacology and neurostimulation therapies for chronic pain have been associated with suboptimal efficacy and limited long-term success, as their mechanisms of action are unclear.
Yet current computational models of pain transmission suffer from several limitations. In particular, dorsal column models do not include the fundamental underlying sensory activity traveling in these nerve fibers.
We developed a (simple) simulation test bed of electrical neurostimulation of myelinated nerve fibers with underlying sensory activity.
This paper reports our findings so far.
Interactions between stimulation-evoked and underlying activities are mainly due to collisions of action potentials and losses of excitability due to the refractory period following an action potential. In addition, intuitively, the reliability of sensory activity decreases as the stimulation frequency increases. 
This first step opens the door to a better understanding of pain transmission and its modulation by neurostimulation therapies.
\end{abstract}

\section{Introduction}

Pain is a protective and adaptive physiological system essential for the survival of many species. However, this system is fragile as damage and malfunction of the nervous system may divert its function, creating a debilitating disease known as chronic pain.
Chronic pain affects about~100~million adults in the~US, with \$560--635~billion in annual medical expenses and lost productivity~\cite{IOM:2011aa}. Chronic pain is primarily treated with neuropharmacology, which may be inadequate or toxic, have negative side effects (\eg{}~addiction to narcotics), and lose efficacy after long-term use~\cite{Breivik:2006aa,Chou:2009aa}. Alternatively, it is also treated with Spinal Cord Stimulation (SCS), an electrical neurostimulation that has potential to reduce the need for drugs and produces less negative side effects. 
However, SCS has been associated with suboptimal efficacy and limited long-term success as their mechanisms of action are unclear~\cite{Turner:2010aa}.

\begin{figure}
	\centering
	\includegraphics{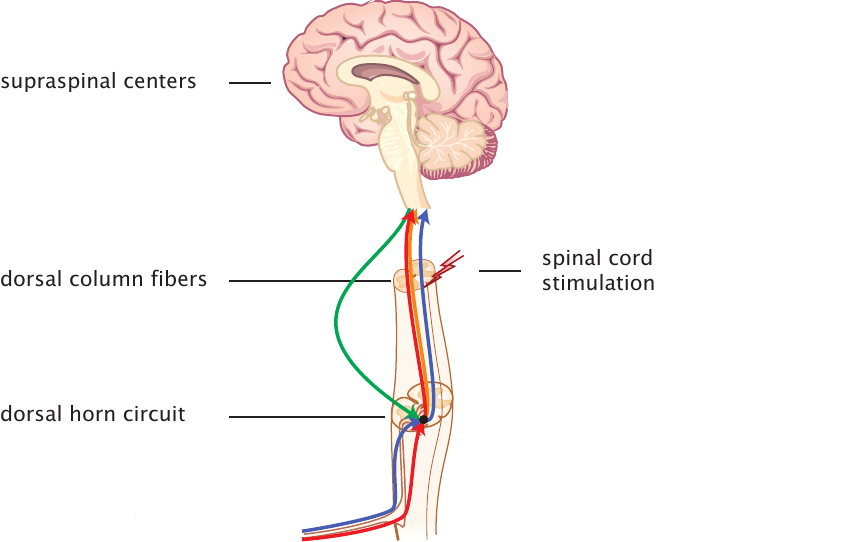}
	\caption{%
	Pain system builds on a tightly regulated dynamical crosstalk between the peripheral nervous system and the brain via the spinal cord.
	Spinal cord stimulation modulates neural activity to suppress chronic pain. 
	}
	\label{fig:pain-system}
\end{figure}

Critical to advancing chronic pain treatment is a deeper mechanistic understanding of pain transmission and modulation under both normal and pathological conditions, which remains largely elusive because the pain system is complex. The pain system builds on a tightly regulated dynamical crosstalk between the peripheral nervous system and the brain via the spinal cord (see~Fig.~\ref{fig:pain-system}). 
Over the past decades, detailed computational models have been used to understand the effects of electrical neurostimulation on dorsal column fibers, a localized but important part of this complex system. In particular, efforts have been made (i) to determine the response of a nerve fiber to an electrical potential field and (ii) to compute the shape of the electrical potential field created by stimulation (see~\cite{Joucla:2012aa} for a recent review on both research directions).  
Although these models reproduce some observed behaviors, none of these models include the fundamental underlying sensory activity (either normal or pathological) traveling in these nerve fibers.

As a first step to address this issue, we developed a simulation test bed of extracellular electrical stimulation of myelinated nerve fibers in the dorsal column \emph{with} underlying sensory activity. 
Unlike previous approaches, our approach considers the potential interactions in dorsal column fibers between neurostimulation-evoked activity \emph{and} underlying sensory activity coming from peripheral nerves. By adding this sensory input, we will gain a better understanding of different mechanisms for SCS analgesia.

The paper is organized as follows.
Section~\ref{sec:methods} describes the computational model of myelinated nerve fibers, the electrical potential field generated by the stimulation, and the underlying afferent activity traveling in these nerve fibers.
Section~\ref{sec:results} reports the results of our simulations: We identify different types of interactions between stimulation and sensory activities and we define a measure of selective relay of sensory neural activities.
Section~\ref{sec:discussion} discusses the physiological interpretation of these results.

\section{Model description} \label{sec:methods}

\begin{figure}
	\centering
	\includegraphics{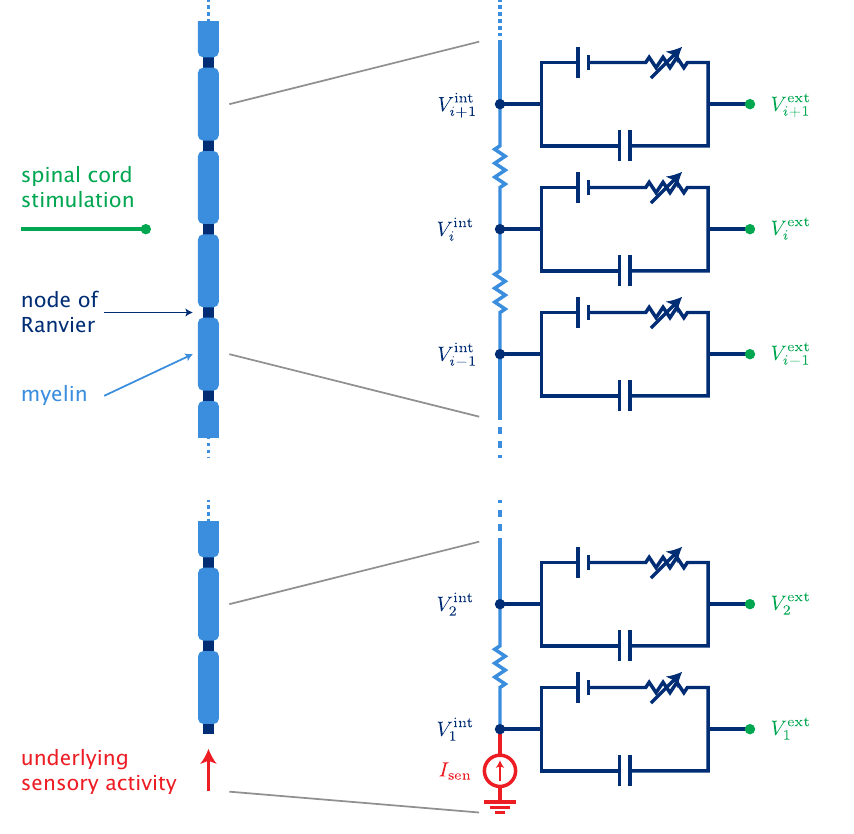}
	\caption{%
		Our model of extracellular electrical stimulation of myelinated nerve fibers in the dorsal column includes the underlying sensory activity as a current source at one end of the nerve fiber.
		}
	\label{fig:fiber-model}
\end{figure}

In this section, we describe our simulation test bed of extracellular electrical stimulation on myelinated nerve fibers in dorsal column \emph{with} underlying sensory activity (see~Fig.~\ref{fig:fiber-model}).
The model is fairly simple: this is both an asset (in the mathematical analysis) and a limitation (in the detailed modeling of more complex structures). As our main interest is to study the effect of interactions between neurostimulation-evoked and underlying activities, this model provides a starting point toward the elucidation of realistic mechanisms.

\subsection{Myelinated nerve fiber}

A myelinated nerve fiber is a cylindrical active membrane (axon), tightly wrapped in an insulating myelin sheath. This myelin sheath is interrupted periodically, leaving short gaps where the axonal membrane is exposed.

Following McNeal's model~\cite{McNeal:1976aa}, a myelinated nerve fiber is represented by an (infinite) series of compartment elements linked by intracellular conductances. 
The dynamics of the membrane potential $V_i = V^{\text{int}}_i - V^{\text{ext}}_i$ at node~$i$ (where $V^{\text{int}}_i$ and $V^{\text{ext}}_i$ are the intracellular and extracellular potentials) read as follows
\begin{equation*}\label{eq:model1}
	\begin{split}
		C_{\text{m}} \,\dot{V}_i + \sum_{k\in\mathcal{K}} \Iion_{i,k} & =  G_{\text{a}} \left(V_{i-1} - 2\,V_i + V_{i+1}\right) \\
		&\quad +  G_{\text{a}}  \left(V^{\text{ext}}_{i-1} - 2\,V^{\text{ext}}_i + V^{\text{ext}}_{i+1}\right) \,,
	\end{split}
\end{equation*}
where  $C_{\text{m}}$ is the membrane capacitance and $G_{\text{a}}$ is the internodal conductance. 
Ionic currents $\Iion_{i,k}$ at node $i$ include a sodium, a fast potassium, and a slow potassium ion channel, as well as a leakage current across the membrane based on the Frankenhaeuser--Huxley model~\cite{Frankenhaeuser:1964aa}, adjusted to experimental data of human sensory fibers at \SI{37}{\celsius} \cite{Schwarz:1995aa}. 
A complete description of the fiber model and its parameters is presented in \cite{Schwarz:1995aa} and \cite{Wesselink:1999aa}.

To numerically compute the response of a \emph{finite} fiber, it is usually assumed that no intracellular axial current flows at the end nodes (`sealed-end' boundary condition)~\cite{Rubinstein:1993aa}. 

\subsection{Electrical potential field generated by the stimulation}

The extracellular medium surrounding a nerve fiber is composed of different regions of the spinal cord (epidural fat, cerebrospinal fluid, white matter, grey matter, etc.), which have different conduction properties~\cite{Struijk:1991aa}.
In addition, the electrode can also take various shapes (single contact, array of contact, etc.) and various configurations (monopolar, bipolar, etc.)~\cite{Medtronic:2007aa}.

However, the extracellular medium may be assumed to be infinite and isotropic with the electrode represented by point sources at the center $x^{\text{c}}_{j}$ of each contact. 
Therefore, the electrical potential field at time~$t$ and position~$x$ is given~by
\begin{equation*}
	\varphi(t,x) = \sum_{j\in\mathcal{C}} \frac{\rho_{\text{m}}}{4\,\pi\,\| x - x^{\text{c}}_{j} \|_2}  \,  I^{\text{sti}}_j(t) \,,
\end{equation*}
where $I^{\text{sti}}_j$ is the current of point source $j$ and $\rho_{\text{m}}$ is the extracellular medium resistivity.
The extracellular potential at node~$i$ is given by $V_i^{\text{ext}}(t) = \varphi(t,x_i)$, where $x_i$ is the position of node $i$.

The stimulation current input $I^{\text{sti}}(t)$ consists of the repetition, at a constant frequency, of symmetrical biphasic pulses with an amplitude of \SI{2.5}{\milli\ampere} and a duration of \SI{350}{\micro\second}. We consider stimulation frequencies ranging from \SIrange[range-units = single]{0}{250}{\hertz}.


\subsection{Underlying sensory activity}

The dorsal column contains nerve fibers that relay peripheral sensory inputs to supraspinal centers.
In normal conditions, these myelinated nerve fibers that originate in the low-threshold primary sensory neurons that mostly signal non-noxious sensory stimuli: proprioception from skeletal muscles and mechanoreception from the skin. 
However, in pathological conditions, mechanical hypersensitivity after injury may also be signaled by abnormal activity in dorsal column fibers~\cite{Baron:2009aa,Song:2012aa}. 
Therefore, the spiking activity in these fibers spans a broad frequency range and exhibits various patterns~\cite{Kajander:1992aa}: regular spike discharge, regular discharge of doublet spikes, bursting patterns, sporadic activity with no regular or predictable firing pattern, etc.

The presence of underlying sensory activity in the nerve fiber is represented by replacing a `sealed-end' boundary condition by a current source at one end of the nerve fiber. Therefore, the dynamics of the first node becomes
\begin{equation*}
	\begin{split}
		C_{\text{m}} \,\dot{V}_1+ \sum_{k\in\mathcal{K}} \Iion_{1,k} & =    G_{\text{a}} \left(V_{2} - V_1\right)  
		  \\
		 &\quad  
		 +  G_{\text{a}}  \left(V^{\text{ext}}_{2} - V^{\text{ext}}_1\right) + I^{\text{sen}}(t) \,,
	\end{split}
\end{equation*}
where the input $I^{\text{sen}}(t)$ represents the underlying activity.

As a first step, the underlying sensory activity input $I^{\text{sen}}(t)$ is modeled as a Poisson train of square pulses with an amplitude of \SI{5}{\nano\ampere} and a duration of \SI{1}{\milli\second}. Therefore, the instantaneous firing rate~$\lambda^{\text{sen}}$ is assumed constant, ranging from \SIrange[range-units = single]{0}{100}{\hertz}.


\begin{figure*}
	\centering
	\includegraphics{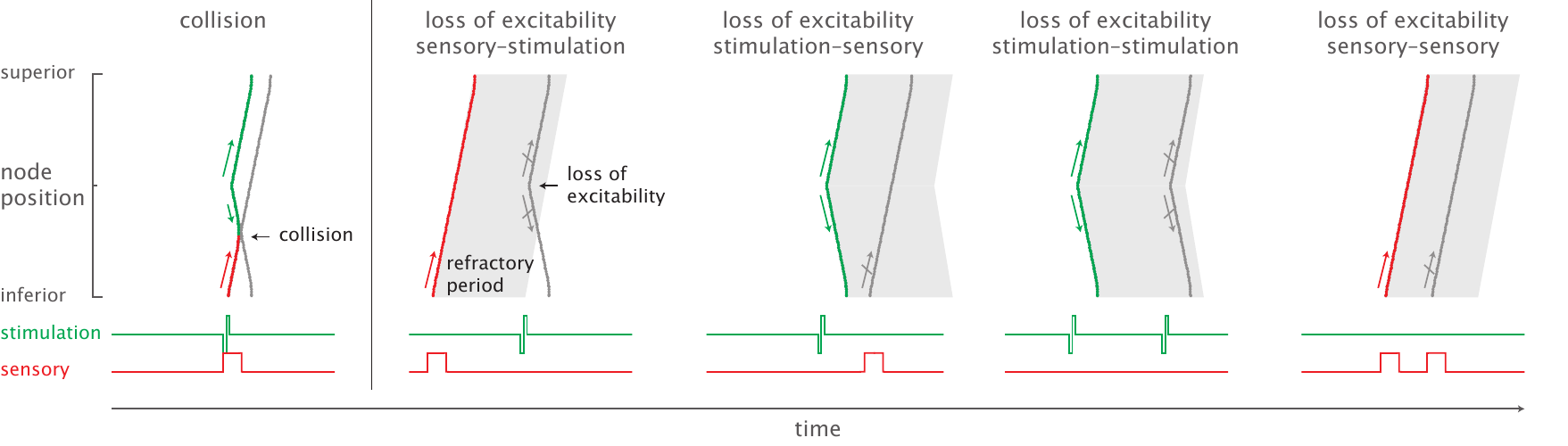}
	\caption{Different interactions between stimulation-evoked activity and sensory activity are illustrated by their scatter plots (horizontal axis is time, vertical axis is position along the nerve fiber). Each dot corresponds to an action potential at a given time $t$ and position $z$ along the fiber. Red and green dots indicate the fiber response to sensory and stimulation inputs, respectively. Sensory action potential waves travel orthodromically from the bottom to the top of the fiber; stimulation-evoked action potential waves travel orthodromically and antidromically from the center to the extremities of the fiber. Gray dots indicate the fiber response in the absence of interactions (collision or loss of excitability), that is, the action potential wave that would be produced by the corresponding input if it was not perturbed by the activity induced by another input.
	}
	\label{fig:interaction}
\end{figure*}

\section{Results} \label{sec:results}

In this section, we show results of our simulation test bed for a monopolar electrode placed at \SI{3.5}{\milli\meter} from the center of a \SI{10}{\centi\meter}-long, \SI{10}{\micro\meter}-large (diameter) fiber.
All the numerical simulations and analyses were performed with MATLAB, MathWorks.

\subsection{Interactions between SCS-evoked and sensory activities}   

We identified different interaction types occurring between simulation-evoked activity and underlying sensory activity.
In Fig.~\ref{fig:interaction}, each panel represents a typical scatter plot for each interaction type. In these panels, a dot is an Action Potential (AP) at node position~$z$ along the fiber and at time~$t$. 
The stimulation input triggers the orthodromic and antidromic propagations of an AP wave (green dots) from the stimulation position at the fiber center toward the fiber ends. 
The sensory input triggers the orthodromic propagation of an AP wave (red dots) from the bottom to the top of the fiber.
Gray dots indicate the fiber response of each input in the absence of the other, that is, artificially without interaction. 

The interaction type depends on the timing of both inputs generating these activities. 


\begin{itemize}
	\item
	A \emph{collision} occurs when the orthodromic sensory AP wave and the antidromic stimulation-evoked AP wave meet and cancel each other. 
	It happens if a sensory pulse is triggered slightly before or after a stimulation pulse, that is, $t^{\text{sen}}_j \in [t^{\text{sti}}_i - \Delta t_{-}^{\text{col}},t^{\text{sti}}_i + \Delta t_{+}^{\text{col}})$.
	\item
	A \emph{loss of excitability sen--sti} occurs when the stimulation input doesn't excite the nerve fiber due to the recent passage of the orthodromic sensory AP wave. 
	It happens if a sensory pulse is triggered before a stimulation pulse, that is, $t^{\text{sen}}_j \in [t^{\text{sti}}_i - \Delta t_{-}^{\text{col}}-\Delta t^{\text{I}},t^{\text{sti}}_i - \Delta t_{-}^{\text{col}})$.
	\item
	A \emph{loss of excitability sti--sen} occurs when the sensory input doesn't excite the nerve fiber due to the recent passage of the antidromic stimulation-evoked AP wave.
	It happens if a sensory pulse is triggered after a stimulation pulse, that is, $t^{\text{sen}}_j \in [t^{\text{sti}}_i + \Delta t_{+}^{\text{col}},t^{\text{sti}}_i + \Delta t_{+}^{\text{col}}+\Delta t^{\text{II}})$.
\end{itemize}

In addition to these interactions, we also identify `self-interactions', that is, interactions between activities generated by the same input.
\begin{itemize}
	\item
	A \emph{loss of excitability sti--sti} occurs when the stimulation input doesn't excite the nerve fiber due to the recent stimulation of the fiber.
	It happens if the frequency of stimulation is too high, that is, $t^{\text{sti}}_{j+1} \in [t^{\text{sti}}_{j} + \Delta t^{\text{III}})$.
	\item
	A \emph{loss of excitability sen--sen} occurs when a sensory input doesn't excite the nerve fiber due to a recent sensory input.
	It happens if two consecutive Poisson pulses happen too quickly, that is, $t^{\text{sen}}_{j+1} \in [t^{\text{sen}}_{j} + \Delta t^{\text{IV}})$.
\end{itemize}

\subsection{Selective relay of sensory neural activities}

Depending on the physiological origin of the underlying sensory activity, we may want to modulate differently the relay of this input with the electrical neurostimulation. 
Typically, we want to block pathological sensory activity (as mechanical hypersensitivity) but we also want to keep relaying normal sensory activity (as proprioception or mechanoreception).
Let us defined reliability as follows
\begin{equation*}
	\text{reliability} = \frac{\text{number of relayed sensory pulses}}{\text{total number of sensory pulses}}\,,
\end{equation*}
where relayed sensory inputs are sensory inputs that travel from the bottom up to the top of the fiber.

\begin{figure}
	\centering
	\includegraphics{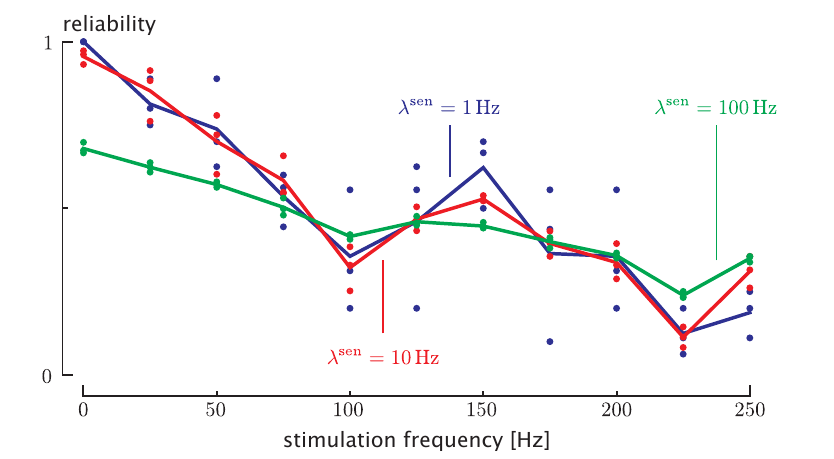}
	\caption{%
		The reliability of sensory input decreases as a function of the stimulation frequency. 
		Dots represents the reliability for 3 different realizations of the Poisson input for each stimulation frequency. The line represents the mean of these 3 reliabilities.
		}
	\label{fig:reliability}
\end{figure}

Figure~\ref{fig:reliability} shows the general decrease in reliability of sensory pulses as a function of the stimulation frequency for different instantaneous firing rates of sensory input. In addition, the simulations show a rebound (localized increase) in reliability around \SI{150}{\Hz} and then a decrease again. We hypothesis that this rebound is due to `loss of excitability sti-sti'. Indeed, \SI{150}{\Hz} corresponds to the frequency at which the nerve fiber doesn't respond anymore to each stimulation pulse due to a too high stimulation frequency.
In addition, Figure~\ref{fig:reliability} shows that a higher reliability is achieved if the sensory input has low instantaneous firing rate.
 
\begin{remark}
The results presented in this subsection are drawn from three simulations of this model with a stochastic input (the sensory activity is Poisson).
These results have to be confirmed by running Monte Carlo simulations, that is, running the same simulation multiple times for different realization of the stochastic input and computing the expected value of the reliability.
\end{remark}



\section{Discussion} \label{sec:discussion}

Dorsal column is the primary target of SCS~\cite{Meyerson:2006aa,Linderoth:2010aa,Guan:2012uq}. 
The dorsal column contains axons that originate in the low-threshold primary sensory neurons that mostly signal non-noxious stimuli. These myelinated axonal fibers relay peripheral sensory inputs to supraspinal centers and have collateral branches projecting to the dorsal horn. 
Since mechanical hypersensitivity after injury may be signaled by abnormal activity in  A$\beta$-fibers~\cite{Baron:2009aa,Song:2012aa} and since the dorsal column contains axons that originate in these neurons, inhibition of A$\beta$-fiber inputs may partially contribute to SCS analgesia, especially for inhibition of mechanical hypersensitivity. 
For example, SCS may reduce activities in dorsal column fibers from reaching second order neurons in brainstem, including cells in the gracile nucleus and cuneate nucleus. 
Thus, SCS may interfere and alter the information coded by physiological sensory inputs. 
Or instead, AP interactions with afferent inputs may also change the pattern of SCS (\eg{}~frequency) and inhibit antidromic APs to activate the segmental spinal network for pain inhibition. 

Our simulation test bed is a first step toward a better understanding of the effect of spinal cord stimulation in relay of sensory input in dorsal column fibers. In particular, we investigated the impact of the stimulation frequency on sensory input with different instantaneous firing rate (frequency content). We identified different types of interactions: collision and losses of excitability. In addition, as expected, a higher stimulation frequency leads to a low reliability of the sensory input. The choice of an optimal stimulation frequency may results from the dual objective to relay a normal sensory input and to block a pathological one, where the instantaneous firing rates of each input are different.

In the future, we plan (i) to consider more complex sensory inputs than Poisson,  such as doublets or bursts, (ii) to derive an analytical expression for the fiber reliability, and (iii) to augment the dorsal column model to include collateral fibers to dorsal horn and dorsal horn circuit itself.

\section*{Acknowledgment}
We would like to thank Dr.~M.~Caterina, Neurosurgery Pain Research Institute, The~Johns Hopkins University School of Medicine, for valuable and insightful discussions.


\bibliographystyle{IEEEtran_without_url}
\bibliography{%
/Users/pierresacre/Work/Research/biblio/biblio_sacre/journal-name-abbrv,%
/Users/pierresacre/Work/Research/biblio/biblio_sacre/bibfile}

\end{document}